\documentstyle[aps,version2,epsf]{revtex}
\newcommand{\qd}{\quad}
\newcommand{\qqd}{\qquad}

\begin{document}
\draft
\begin{title}
Noise-induced switching between vortex states with different
polarization in classical two-dimensional easy-plane magnets
\end{title}
\author{Yuri Gaididei}
\begin{instit} 
Institute for Theoretical Physics, 252143 Kiev, Ukraine
\end{instit}
\author{Till Kamppeter and Franz G. Mertens}
\begin{instit}
Physikalisches Institut, Universit\"at Bayreuth, D-95440 Bayreuth,
Germany
\end{instit}
\author{Alan Bishop}
\begin{instit}
Theoretical Division and Center for Nonlinear Studies, Los Alamos
National Laboratory, Los Alamos, New Mexico 87545, USA
\end{instit}

\begin{abstract}
  In the 2-dimensional anisotropic classical Heisenberg model with
  $XY$-symmetry there are nonplanar vortices which exhibit a
  localized structure of the $z$-components of the spins around the
  vortex center. We study how thermal noise induces a transition of
  this structure from one polarization to the opposite one. We
  describe the vortex core by a discrete Hamiltonian and consider a
  stationary solution of the Fokker-Planck equation. We find a bimodal
  distribution function and calculate the transition rate using
  Langer's instanton theory (1969). The result is compared with
  Langevin dynamics simulations for the full many-spin model.
\end{abstract}
\pagebreak
%
%
\section{Introduction}
There are several classes of quasi-2D (two-dimensional) magnetic
materials for which the ratio of inter- to intraplane magnetic
coupling constants is typically $10^{-3} - 10^{-6}$: (1) Layered
magnets \cite{deJongh90,Regnault89,Regnault91,Hirakawa82}, like
K$_2$CuF$_4$, Rb$_2$CrCl$_4$, (CH$_3$NH$_3$)$_2$CuCl$_4$, and
BaM$_2$(XO$_4$) with M$=$Co, Ni, ... and X$=$As, P, ...  (2) CoCl$_2$
graphite intercalation compounds \cite{Wiesler89}, (3) magnetic lipid
layers, like manganese stearate \cite{Pom88}. Many of these materials
can be described by the classical 2D Heisenberg model with $XY$- or
''easy-plane'' symmetry (section II).

In this model vortices play the decisive role: they are responsible
for a topological phase transition \cite{iida63,gard} at the
Kosterlitz-Thouless temperature $T_c$ and, above $T_c$, for ''central
peaks'' in the dynamic form factors for the spin correlations. The
central peaks were observed in inelastic neutron scattering
experiments \cite{Hutch86,Regnault86,Regnault87,Bramwell88,Wiesler94}
and in combined Monte Carlo/Spin Dynamics simulations
\cite{fgm87,fgm89,Gouvea89,Voelkel91b,Voelkel92,Voelkel93}. The
observed central peaks agree qualitatively, partially even
quantitatively, with the central peaks which were obtained by a
vortex-gas approach
\cite{fgm87,fgm89,Gouvea89,Voelkel91b,Voelkel92,Voelkel93}.

There are two types of static vortex solutions whose structure and
energy differ, depending on the anisotropy of the Heisenberg exchange
interaction \cite{Gouvea89}. For strong anisotropy (i. e., if the
anisotropy parameter $\delta$ exceeds a threshold $\delta_c$) only
planar vortices are stable for which all spins are lying in the easy
plane ($xy$-plane). For weak anisotropy $(0<\delta <\delta_c)$ only
nonplanar vortices are stable, which exhibit a localized structure of
the $z$-components of the spins around the vortex center. In addition
to the vorticity $q = \pm1, \, \pm 2, ...$, the nonplanar vortices
have a second topological charge $p$. It is denoted ''polarization''
because its sign determines the side of the $xy-$plane to which the
out-of-plane vortex structure points. The planar vortices can be
considered as having $p = 0$.

The product $qp$ of the topological charges determines the dynamics
because the vortices are subject to a ''gyrocoupling force'' $\vec{G}
\times \vec{V}$, which is formally equivalent to the Lorentz force
\cite{thiele73,huber82}: $\vec{V}$ is the velocity of the vortex
center, but instead of an external magnetic field we have here an
intrinsic quantity, produced by the vortex itself and carried along
with it: The ''gyrovector'' $\vec{G} = 2 \pi qp \vec{{\rm e}}_z$ which
is orthogonal to the $xy$-plane. The formula for $\vec{G}$ was derived
in the continuum limit and, strictly speaking, $\vec{G}$ is conserved
only in this limit. Nevertheless, spin dynamics simulations for 1 or 2
vortices showed that the direction of $\vec{G}$ (or the sign of $p$,
because $q$ is always conserved) does not change during the simulation
\cite{voelkel91a,Mertens94}.

However, we know so far three exceptions, i. e. situations in which
the out-of-plane vortex structure can suddenly make a transition from
one polarization to the opposite one. As the direction of $\vec{G}$ is
reversed, this has a drastic effect on the dynamics: The direction of
the gyrocoupling force is also reversed, which means that the
direction of the vortex motion is
reversed, too. The three transition mechanisms are:\\

(1) \underline{Interaction with spin waves.} The easiest way to see
this is to use ''dirty'' initial conditions for the spin dynamics
simulation \cite{Schnitzer96}: E. g., a structure which is not a good
approximation to the $1$-vortex solution (this solution can be
obtained numerically by an iteration procedure \cite{Schnitzer96}).
Then many spin waves are radiated at the beginning of the simulation,
while the approximate vortex structure adapts to the lattice and
becomes a ''good'' solution (numerically identical to the above
solution obtained by iterations). The emitted spin waves form a magnon
gas; i. e. the vortex moves in a kind of magnon thermostat and
transitions to the opposite polarization occur with a certain
probability which depends on how dirty
the initial condition was.\\

(2) \underline{An $ac$ magnetic field.} If the amplitude of a field
which rotates in the easy plane is larger than a threshold value, a
transition to the opposite polarization occurs. In contrast to (1),
the reverse process does not occur because the field breaks the
symmetry of the two
polarizations. This will be the subject of a forthcoming paper.\\

(3) \underline{Thermal noise.} This has some similarity with (1),
although that is a deterministic zero-temperature effect. In section
II we implement white noise into the microscopic equations (the
Landau-Lifshitz eq.) by adding stochastic magnetic fields to the local
fields in which every spin precesses. In this way we model the
interactions of the spin degrees of freedom with thermostat degrees of
freedom (magnons, phonons etc.). We consider a stationary solution
${\cal P}_{st}$ of the Fokker-Planck equation, using a reduced
Hamiltonian which models the vortex core. Such a core Hamiltonian was
used in \cite{wysin94,wysin97} for the calculation of $\delta_c$. For
a certain parameter range, ${\cal P}_{st}$ exhibits two maxima (for
the two possible polarizations of a nonplanar vortex) and a saddle
point (corresponding to the planar vortex).

In section III we calculate the probability flux over the region
around the saddle point using Langer's instanton theory
\cite{langer69}. Here we use the fact that for $\delta \rightarrow
\delta_c$ there is a soft mode among the normal modes which were
obtained numerically for a system with one vortex \cite{wysin95}.

Finally, our prediction for the transition rate is tested by Langevin
dynamics simulations, i. e. by integration of the stochastic
Landau-Lifshitz equation.For these tests the design of the
simulations, including the choice of the parameter ranges, turns out
to be decisive.

\section{Hamiltonian and the Fokker-Planck equation}
We consider a Heisenberg model with XY- or easy-plane symmetry with
  classical spins $\vec{S}_{\vec{n}}$
  located on the sites $\vec{n}=(n_x,n_y)$ of a square lattice
\begin{equation}
\label{eq1}
H=-\sum_{\vec{n},\vec{\Delta}}J_{\vec{\Delta}}\,\left(S^x_{\vec{n}}\,
S^x_{\vec{n}-\vec{\Delta}}+
S^y_{\vec{n}}\,S^y_{\vec{n}-\vec{\Delta}}+\lambda\,S^z_{\vec
{n}}\,S^z_{\vec{n}-\vec{\Delta}}\right)
\end{equation}
where $\lambda$ is the anisotropy parameter ($0\leq\lambda<\,1$), 
$J_{\vec{\Delta}}\equiv\, J$ is the exchange integral and $\vec{\Delta}=
(\Delta_x,\Delta_y)$ is a vector which connects a spin with its nearest 
neighbors ($\Delta_x=\pm 1,\Delta_y=0$ or $\Delta_y=\pm 1,\Delta_x=0$).
The spin dynamics is governed by the Landau-Lifshitz equation. Since we
want to study the interaction with thermal noise, we implement a noise
and a damping term
\begin{eqnarray}
\label{eq3}
\frac{d}{dt}\vec{S}_{\vec{n}}=-\vec{S}_{\vec{n}}\times\left(\frac{\partial
H}{\partial\vec{S}_{\vec{n}}}+\vec{h}_{\vec{n}}(t)\right)+\gamma\,
\vec{S}_{\vec{n}}\times\left(\vec{S}_{\vec{n}}\times\frac{\partial
H}{\partial\vec{S}_{\vec{n}}}\right)
\end{eqnarray}
Here we have added a stochastic magnetic field $\vec{h}_{\vec{n}}(t)$ to
the local field $\frac{\partial H}{\partial\vec{S}_{\vec{n}}}$ in
which the spin $\vec{S}_{\vec{n}}$ precesses. Since $\vec{h}_{\vec{n}}$ is
multiplied with $\vec{S}_{\vec{n}}$, this means multiplicative noise.

Another way to obtain the stochastic term in Eq. (\ref{eq3}) consists
in adding to the Hamiltonian interactions between the spins and local
stochastic magnetic fields
\begin{equation}
\label{stoch}
V(t)=-\sum_{\vec{n}}
\vec{h}_{\vec{n}}(t)\,\vec{S}_{\vec{n}}\, .
\end{equation}
We use Gaussian white noise with
\begin{eqnarray}
\label{eq2}
<h^{\alpha}_{\vec{n}}(t)>&=&0\enspace,\nonumber\\
<h^{\alpha}_{\vec{n}}(t)\,h^{\alpha'}_{\vec{n}'}(t')>&=&2\,D_{\alpha}\,
\delta_{\vec{n}\vec{n}'}\,\,\delta_{\alpha \,\alpha'}\,\delta(t-t')\, ,
\end{eqnarray}
where $D_{\alpha}$ is the variance of the noise. In order to preserve
the isotropy in the easy plane we demand
\begin{eqnarray}
\label{var}
D_x=D_y \equiv D \enspace.
\end{eqnarray}
The last term in Eq. (\ref{eq3}) represents damping in the
Landau-Lifshitz form (see \cite{iida63}). An alternative would be the
Gilbert damping which yields the same results, however, as we will use
only very small damping coefficients.

It is convenient to use a representation for the classical
 spin vector $\vec{S}_{\vec{n}}$ in terms of two angles of rotation
 $\theta_{\vec{n}}$ and $\Phi_{\vec{n}}$
\begin{eqnarray}
\label{eq4}
\vec{S}_{\vec{n}}=S\{\sin\theta_{\vec{n}}\cos\Phi_{\vec{n}},
\sin\theta_{\vec{n}}\sin\Phi_{\vec{n}},\cos\theta_{\vec{n}}\}
\end{eqnarray}
The variables $M_{\vec{n}}=\cos\theta_{\vec{n}}$ and $\Phi_{\vec{n}}$
constitute a pair of canonically conjugated variables, which means
that in the no-damping case ($\gamma=0$)
\begin{eqnarray}
\label{eq5}
\frac{d \Phi_{\vec{n}}}{dt}&=&\frac{\partial(H+V)}{\partial
M_{\vec{n}}}\enspace,\nonumber\\
\frac{d M_{\vec{n}}}{dt}&=&-\frac{\partial (H+V)}{\partial
  \Phi_{\vec{n}}}\enspace.
\end{eqnarray}
Here
\begin{eqnarray}
\label{eq6}
H&=&-J\sum_{\vec{n},\vec{\Delta}}\left(\lambda\,M_{\vec{n}}\,
M_{\vec{n}-\vec{\Delta}}+P_{\vec{n}}\,P_{\vec{n}-\vec{\Delta}}\,
\cos(\Phi_{\vec{n}}-\Phi_{\vec{n}-\vec{\Delta}})\right)\nonumber\\
V(t)&=&-\sum_{\vec{n}}\left(
\,h^z_{\vec{n}}(t)\,M_{\vec{n}}+P_{\vec{n}}\,
(h^x_{\vec{n}}(t)\,\cos(\Phi_{\vec{n}})+
h^y_{\vec{n}}(t)\,\sin(\Phi_{\vec{n}}))\right)
\end{eqnarray}
is the Hamiltonian of the system in terms of the new variables and 
 $P_{\vec{n}}=\sqrt{1-M^2_{\vec{n}}}$.

Using the variables $M_{\vec{n}}$ and  $\Phi_{\vec{n}}$ the
Landau-Lifshitz equation (\ref{eq3}) can be written as a set
of coupled stochastic equations
\begin{eqnarray}
\label{eq7}
\frac{d \Phi_{\vec{n}}}{dt}&=&\frac{\partial H}{\partial M_{\vec{n}}}-
\frac{\gamma}{1-M^2_{\vec{n}}}\frac{\partial H}{\partial
\Phi_{\vec{n}}}\,+\,
f_{\vec{n}}\left(M_{\vec{n}},\Phi_{\vec{n}},t \right) \enspace,
\nonumber\\
\frac{d M_{\vec{n}}}{dt}&=&-\frac{\partial H}{\partial \Phi_{\vec{n}}}-
\gamma\,(1-M^2_{\vec{n}})\,\frac{\partial H}{\partial M_{\vec{n}}}
\,+g_{\vec{n}}\left(M_{\vec{n}},\Phi_{\vec{n}},t \right) \enspace.
\end{eqnarray}
where
\begin{eqnarray}
\label{eq8}
f_{\vec{n}}\left(M_{\vec{n}},\Phi_{\vec{n}},t \right)&=&-h^z_{\vec{n}}(t)+
\frac{M_{\vec{n}}}{P_{\vec{n}}}\,\left(h^x_{\vec{n}}(t)\,\cos(\Phi_{\vec{n}})+
h^y_{\vec{n}}(t)\,\sin(\Phi_{\vec{n}})\right)\enspace,\nonumber\\
g_{\vec{n}}\left(M_{\vec{n}},\Phi_{\vec{n}},t
\right)&=&-P_{\vec{n}}\,\left(h^x_{\vec{n}}(t)\,\sin(\Phi_{\vec
{n}})-
h^y_{\vec{n}}(t)\,\cos(\Phi_{\vec{n}})\right)
\end{eqnarray}
are multiplicative stochastic forces. From Eqs (\ref{eq2}), (\ref{var})
and (\ref{eq8}) we obtain
\begin{eqnarray}
\label{eq9}
<f_{\vec{n}}\left(M_{\vec{n}},\Phi_{\vec{n}},t \right)\,
f_{\vec{n}'}\left(M'_{\vec{n}'},\Phi'_{\vec{n}'},t'
\right)>\,&=&
\,2\,\delta(t-t')\,\delta_{\vec{n}\vec{n}'}\,\left(D_z+D\,
\frac{M_{\vec{n}}\,M'_{\vec{n}}}
{P_{\vec{n}}P'_{\vec{n}}}\cos(\Phi_{\vec{n}}-
\Phi'_{\vec{n}})\right),\nonumber\\
<g_{\vec{n}}\left(M_{\vec{n}},\Phi_{\vec{n}},t \right)\,
g_{\vec{n}'}\left(M'_{\vec{n}'},\Phi'_{\vec{n}'},t'
\right)>\,&=&\,2\,D\,\delta(t-t')\,\delta_{\vec{n}\vec{n}'}\,
P_{\vec{n}}P'_{\vec{n}}\cos(\Phi_{\vec{n}}-
\Phi'_{\vec{n}})~~~~\nonumber\\
<f_{\vec{n}}\left(M_{\vec{n}},\Phi_{\vec{n}},t \right)\,
g_{\vec{n}'}\left(M'_{\vec{n}'},\Phi'_{\vec{n}'},t'
\right)>\,&=&
\,2\,D\,\delta(t-t')\,\delta_{\vec{n}\vec{n}'}\,M_{\vec{n}}
\frac{P'_{\vec{n}}}{P_{\vec{n}}}\sin(\Phi_{\vec{n}}-
\Phi'_{\vec{n}})~~~~~
\end{eqnarray}

We have introduced the stochastic magnetic fields
$\vec{h}_{\vec{n}}(t)$ to model the interaction of the spin degrees of
freedom with thermostat degrees of freedom: phonons, electrons, other
magnetic excitations etc. However, it is clear that the thermostat
excitations are characterized by finite correlation times and a more
appropriate modelling of the influence of these excitations would be
to use colored noise. The problem under consideration is very
complicated, however, if considered in the framework of a
nonwhite-noise approach. Therefore we restrict ourselves to the case
(\ref{eq2}), where we understand the white noise approach as a
limiting case of the colored-noise process and therefore we consider
Eqs (\ref{eq7}), (\ref{eq8}) as Stratonovich stochastic differential
equations \cite{gard}.

From Eqs (\ref{eq7})-(\ref{eq9}) we obtain that the equation for the
probability density function
\begin{eqnarray}
\label{eq10}
{\cal P}(m_{\vec{n}},\phi_{\vec{n}},t)=
\langle\prod_{\vec{n}}\,\delta
\left(m_{\vec{n}}-M_{\vec{n}}(t)\right)\delta \left(
\phi_{\vec{n}}-\Phi_{\vec{n}}(t)\right)\rangle
\end{eqnarray}
has the form
\begin{eqnarray}
\label{eq11}
\frac{\partial}{\partial t}{\cal P}&=&
\sum_{\vec{n}}\frac{\partial}{\partial \phi_{\vec{n}}}\left(\,(
-\frac{\partial\,H}{\partial\, m_{\vec{n}}}\,+
\frac{\gamma}{1-m^2_{\vec{n}}}
\frac{\partial\,H}{\partial \,\phi_{\vec{n}}}){\cal P}+(D_z\,+
\,D\,\frac{m^2_{\vec{n}}}{1-m^2_{\vec{n}}}\,)
\frac{\partial}{\partial \,\phi_{\vec{n}}}{\cal P}\right)-
\nonumber\\
&&\sum_{\vec{n}}\frac{\partial}{\partial m_{\vec{n}}}\left(\,(
\frac{\partial\,H}{\partial\, \phi_{\vec{n}}}\,+
\gamma\,(1-m^2_{\vec{n}})
\frac{\partial\,H}{\partial \,m_{\vec{n}}}){\cal P}+
\,D\,(1-m^2_{\vec{n}})\,
\frac{\partial}{\partial \,m_{\vec{n}}}{\cal P}\right)
\end{eqnarray}
As was mentioned above, the stochastic magnetic fields
$\vec{h}_{\vec{n}}(t)$ model the interaction with thermostat degrees
of freedom. Therefore it is quite natural to demand that Eq.
(\ref{eq11}) has a stationary solution in the form of the Gibbs
distribution
\begin{eqnarray}
\label{eq12}
{\cal P}_{st}\sim\exp\left(-\frac{H}{T}\right).
\end{eqnarray}
It is seen from Eq. (\ref{eq11}) that the function (\ref{eq12}) is a
steady-state solution of the Fokker-Planck equation (\ref{eq11})
when the fluctuation-dissipation condition
\begin{eqnarray}
\label{eq13}
D_z=D=\gamma\,T
\end{eqnarray}
is fulfilled. Here $T$ is the temperature of the crystal. Under the
condition (\ref{eq13})
the Fokker-Planck equation for the  function ${\cal P}$ has
the form
\begin{eqnarray}
\label{eq14}
\frac{\partial}{\partial t}{\cal P}&=&
\sum_{\vec{n}}\frac{\partial}{\partial \phi_{\vec{n}}}\left(\,(
-\frac{\partial\,H}{\partial\, m_{\vec{n}}}\,+
\frac{\gamma}{1-m^2_{\vec{n}}}
\frac{\partial\,H}{\partial \,\phi_{\vec{n}}}){\cal P}+
\,\gamma\,T\,\frac{1}{1-m^2_{\vec{n}}}\,
\frac{\partial}{\partial \,\phi_{\vec{n}}}{\cal P}\right)-
\nonumber\\
&&\sum_{\vec{n}}\frac{\partial}{\partial m_{\vec{n}}}\left(\,(
\frac{\partial\,H}{\partial\, \phi_{\vec{n}}}\,+
\gamma\,(1-m^2_{\vec{n}})
\frac{\partial\,H}{\partial \,m_{\vec{n}}}){\cal P}+
\,\gamma\,T\,(1-m^2_{\vec{n}})\,)
\frac{\partial}{\partial \,m_{\vec{n}}}{\cal P}\right)
\end{eqnarray}

Let us consider first the equilibrium properties of the system.  We
assume that a vortex is situated in the center of a unit cell at the
origin of a coordinate system. The static in-plane vortex
($m_{\vec{n}}=0$) is characterized by the angles $\Phi^0_{\vec{n}}$
which satisfy the equation
\begin{eqnarray}
\label{eq15}
\sum_{\vec{\Delta}}\sin(\Phi^0_{\vec{n}}-
\Phi^0_{\vec{n}-\vec{\Delta}})=0.
\end{eqnarray}
The $\Phi^0_{\vec{n}}$ are approximately given by the usual in-plane
vortex structure
\begin{eqnarray}
\label{eq16}
\Phi^0_{\vec{n}}=\,q\,\arctan\left(\frac{n_y}{n_x}\right)
\end{eqnarray}
where $n_x,n_y=(2n+1)/2,n=0,\pm1,\pm2,..$ (the lattice constant is
chosen equal to 1) and the integer $q$ is the vorticity. It is known
\cite{wysin94}, \cite{wysin95} that the in-plane vortex is stable for
$0\,<\,\lambda\,<\,\lambda_c$ where the critical value $\lambda_c$ of
the anisotropy parameter depends on the type of the lattice (e.g.  for
square lattices $\lambda_c \simeq\, 0.703$ \cite{wysin97}).  For
$\lambda\,>\,\lambda_c$ the in-plane vortex becomes unstable and an
out-of-plane vortex is created.  To gain insight as to how the
temperature influences the stability conditions we need a reduced form
of the Hamiltonian (\ref{eq6}) which effectively takes into account
both types of vortices: in-plane and out-of-plane.  Such an effective
Hamiltonian was proposed in \cite{wysin94}.  It was shown in
\cite{wysin94} that the dynamics of the vortex instability can be
understood under the following assumptions:
 
 i) The in-plane angles $\Phi^0_{\vec{n}}$ for static in-plane and
 out-of-plane vortices are given by Eq. (\ref{eq15}).
 
 ii) The deviations $\psi_{\vec{n}}=\Phi_{\vec{n}}-\Phi^0_{\vec{n}}$
 of the in-plane angles from their static values are radially
 symmetric. They strongly decay with the distance 
 $r_{\vec{n}}=\sqrt{(n_x-1/2)^2+(n_y-1/2)^2}$
  from the vortex center: 
  \begin{equation}
\psi_{\vec{n}}= \left\{ \begin{array}{rl}
                      \psi_1 \; , & \mbox{for $\vec{n}=\pm(1/2,1/2),  
                     \pm(-1/2,1/2)$} \\
                  \displaystyle{\psi_2} \;
                      , & \mbox{for $\vec{n}=\pm(3/2,1/2), \pm(-1/2,3/2) $}\\
                      \displaystyle{\psi_3} \;
                      , & \mbox{for $\vec{n}=\pm(1/2,3/2), \pm(-3/2,1/2) $}\\
                      \displaystyle{0} \;
                      , & \mbox{otherwise}
                   \end{array}
              \right.
\end{equation}

iii) The deviations of the out-of-plane components are also radially
 symmetric and decay strongly
 \begin{equation}
m_{\vec{n}}= \left\{ \begin{array}{rl}
                      m_1 \; , & \mbox{for $\vec{n}=\pm(1/2,1/2),  
                     \pm(-1/2,1/2)$} \\
                  \displaystyle{m_2} \;
                      , & \mbox{for $\vec{n}=\pm(3/2,1/2), \pm(-1/2,3/2) $}\\
                      \displaystyle{m_3} \;
                      , & \mbox{for $\vec{n}=\pm(1/2,3/2), \pm(-3/2,1/2) $}\\
                      \displaystyle{0} \;
                      , & \mbox{otherwise}
                   \end{array}
              \right.
\end{equation}
 
Under these assumptions the dynamics of the vortex core is described by
the following Hamiltonian
\begin{eqnarray}
\label{eq17}
H_c&=&-4\,J\,\{\lambda\,\left(m^2_1+m_1\,(m_2+m_3)+m_2\,m_3\right)
\nonumber\\
&&+\,\cos\delta_1\,p_{1}\,\left(p_{2}\,\cos(\psi_1-\psi_2)+
p_{3}\,\cos(\psi_1-\psi_3)\right)
\nonumber\\
&&+\,(\cos\delta_1\,+\cos\delta_2)\left(p_{2}\,
\cos \psi_2+
p_{3}\,\cos \psi_3\right)\nonumber\\
&&+\,p_{2}p_{3}\,
\sin(2\delta_1)\cos(\psi_2-\psi_3)\}
\end{eqnarray}
with 
\begin{eqnarray}
\label{eq21}
p_{\alpha}=\sqrt{1-m^2_{\alpha}},~~\alpha=1,2,3
\end{eqnarray}
and
\begin{eqnarray}
\label{eq18}
\delta_1=\Phi^0_{1/2,1/2}- \Phi^0_{3/2,1/2}\enspace,~~~
\delta_2=\Phi^0_{3/2,1/2}- \Phi^0_{5/2,1/2}\enspace.
\end{eqnarray}
Using the approximation $\psi_2=\psi_3,m_2=m_3$ 
and the static in-plane angle distribution 
(\ref{eq16}), we get $\cos\delta_1=2/\sqrt{5}, \cos\delta_2=8/\sqrt{65}$
and in this case the Hamiltonian (\ref{eq17}) coincides with the
Hamiltonian in \cite{wysin94}.

Being interested in the distribution of the out-of-plane components
$m_\alpha$ we integrate the function (\ref{eq12}) with respect to the
in-plane angles $\psi_\alpha$. We obtain a reduced stationary
probability density
\begin{eqnarray}
\label{eq20}
\lefteqn{{\cal P}_{st}(m_1,m_2,m_3)=\frac{1}{{\cal N}}\,
e^{\lambda\beta\left(m^2_1+m_1\,(m_2+m_3)+m_2\,m_3\right)}} \nonumber\\
 & & \qqd \int\limits_{0}^{2\pi}\,d\phi\,
e^{\beta\sin(2\delta_1)p_2\,p_3\cos\phi}
I_{0}(\beta\cos\delta_1\,p_1\sqrt{p^2_2+p^2_3+2p_2p_3\cos\phi})\times 
\nonumber\\
 & & \qqd I_{0}(\beta(\cos\delta_1+\cos\delta_2)\sqrt{p^2_2+p^2_3+2p_2p_3
\cos\phi}) \enspace,
\end{eqnarray}
where $\beta=4\,J/T$ is a dimensionless inverse temperature. 
${\cal N}$ is the normalization factor and $I_{0}(x)$ is a modified 
Bessel function \cite{HighTransFunc53,AbSt72}.

The analysis of the function (\ref{eq20}) shows that it has a unique
maximum at $m_1=m_2=m_3=0$ if the anisotropy parameter $\lambda$ is
below a temperature dependent threshold value $\lambda_c(T)$.  This
case corresponds to the stable in-plane vortex. If
\begin{eqnarray}
\label{eq23}
\lambda\,>\,\lambda_c(T)
\end{eqnarray}
the function (\ref{eq20}) has two maxima at $m_1=\pm\,m^0_1,
m_2=\pm\,m^0_2, m_3=\pm\,m^0_3$ and a saddle point at $m_1=m_2=m_3=0$.
In this case the probability density function (\ref{eq20}) describes a
bistable system of two out-of-plane vortex structures with opposite
polarizations.

Let us illustrate this statement by a crude approach when only two
degrees of freedom $m_1$ and $\psi_1$ are included. In this case the
core Hamiltonian (\ref{eq17}) simplifies to
\begin{eqnarray}
\label{core}
H_c=-4\,J\{\lambda\,m^2_1+2\cos\delta_1 p_{1}\cos\psi_1\}
\end{eqnarray}
The corresponding stationary distribution ${\cal P}_{st}(m_1,\psi_1)$
is plotted in Fig. 1a (without normalization).
The reduced stationary probablity density (\ref{eq20}) has the form
\begin{eqnarray}
\label{red}
{\cal P}_{st}(m_1)&=&
\exp\left(\beta\lambda\,m^2_1\right)\,I_0\left(2\beta\cos\delta_1 p_{1}\right)
/{\cal N}\enspace,\nonumber\\
{\cal N}&=&\int\limits_{-1}^{1}\,d\,m_1\,
\exp\left(\beta\lambda\,m^2_1\right)\,I_0\left(2\beta\cos\delta_1 p_{1}\right)
\end{eqnarray}
The function ${\cal P}_{st}(m_1)$ 
describes a bimodal distribution  (see Fig.1(b)) in the range
\begin{eqnarray}
\label{bimod}
\beta\,\cos^2\delta_1\,>\,\lambda\,>\,\cos\delta_1
\frac{I_1(2\beta\cos\delta_1)}{I_0(2\beta\cos\delta_1)}
\end{eqnarray}
A more accurate approach is based on the expansion of the Hamiltonian
(\ref{eq17}) into a series with respect to $\{\psi\}$.
Then in the harmonic approximation with respect to the $\psi_\alpha$
the stationary probability density is determined by the
expression
\begin{eqnarray}
\label{229}
&&{\cal P}_{st}(m_1,m_2,m_3)= \nonumber\\
&&\qqd\frac{1}{{\cal N}}\,
\frac{\exp\left(\beta\,H_c(\{m\},\{\psi\}=0)\right)}
{\sqrt{p_1\,p_2\,p_3\,(p_2+p_3)
\,(\sin(2\delta_1)\,(p_2+p_3)+p_1\cos\delta_1+\cos\delta_1+\cos\delta_2
)}}
\end{eqnarray}
The function (\ref{229}) describes a bimodal distribution if
\begin{eqnarray}
\label{eq230}
\lambda\,>\,\lambda(\beta)\equiv \frac{4 \beta \cos\delta_1 (2 \cos\delta_1+
\cos\delta_2+2\sin 2\delta_1)-2\sin 2\delta_1-\cos\delta_2-3\cos\delta_1}
{4 \beta (2\sin 2\delta_1+\cos\delta_2+2\cos\delta_1)}\enspace.
\end{eqnarray}
Thus the function (\ref{eq20}) has two maxima at $m_1=\pm\,m^0_1, 
m_2=\pm\,m^0_2, m_3=\pm\,m^0_3$ and a saddle point at $m_1=m_2=m_3=0$. 
The phase diagram (the bifurcation curve $\lambda(\beta)$) is shown in
Fig. 2. It is worth noting that for a
given anisotropy parameter $\lambda$ the phase which corresponds to the
in-plane vortex is always the low-temperature phase.

\section{Switching rate}

Following Langer \cite{langer69} (see also Ref.~\cite{htb90}) it is
convenient to introduce a new set of variables
$\{\eta\}=(\eta_1,...\eta_{2 N})$ which consists of $N$ out-of-plane
spin deviations $(\eta_1,...\eta_{ N})= \{m_{\vec{n}}\}$ and $N$
canonically conjugated variables $(\eta_{N+1},...\eta_{2
  N})=\{\phi_{\vec{n}}\}$ and to write the Fokker-Planck equation
(\ref{eq14}) in the form
\begin{eqnarray}
\label{eq24}
\frac{\partial {\cal P}(\{\eta\},t)}{\partial
t}=\sum_{i,j}\,\frac{\partial}{\partial\eta_i}M_{i,j}\left(\frac{\partial
E}{\partial \eta_j}{\cal P}+T
\frac{\partial }{\partial \eta_j}{\cal P}\right)
\end{eqnarray}
where
\begin{eqnarray}
\label{eq24a}
M_{i,j}=\Gamma_i\delta_{ij}-A_{i,j}
\end{eqnarray}
with 
\begin{equation}
\label{gamma}
\Gamma_i= \left\{ \begin{array}{rl}
                      \frac{\gamma}{1-\eta^2_{i}}  \; , 
                      & \mbox{for $i\,\leq \,N$} \\
                     \displaystyle{\gamma (1-\eta^2_{i})} \;
                      , & \mbox{for $i\,\geq \,N+1$}
                   \end{array}
              \right.
\end{equation}
and  $A_{i,j}$ is the following antisymmetric matrix:
\begin{equation}
\label{anti}
A_{i,j}= \left\{ \begin{array}{rl}
                      \delta_{i+N,j} \; , & \mbox{$i\,\leq \,N$} \\
                  \displaystyle{-\delta_{i,j+N}} \;
                      , & \mbox{$j\,\leq \,N$}\\
                      \displaystyle{0} \;
                      , & \mbox{otherwise}
                   \end{array}
              \right.
\end{equation}
$E(\{\eta\})$ is the Hamiltonian of the system expressed in terms of 
the variables $\{\eta\}$.

We are interested in a switching process between the vortex states
with different polarization. Therefore we consider the
anisotropy-temperature region (see Fig.\ 2) where the out-of-plane
vortices are stable.  In this case the energy function $E(\{\eta\})$
has a locally stable state at $\{\eta_0\}$ (an out-of-plane vortex
with positive polarization) which is separated by an energy barrier
from another stable state $\{-\eta_0\}$ (an out-of-plane vortex with
negative polarization). We assume that the system is initially
prepared in a vortex state with, say, positive polarization, and we
consider the relaxation process as an escape process from the
potential well which corresponds to the vortex $\{\eta_0\}$ neglecting
the backward process. Another possibility to make the vortices with
different polarization non-equivalent is to apply a constant magnetic
field oriented along the hard-axis (perpendicular to the easy-plane).
The in-plane vortex $\{\bar{\eta}\}$ with the same vorticity as the
out-of-plane vortex corresponds to the energy barrier which must be
overcome. The point $\{\bar{\eta}\}$ is a saddle point of
$E(\{\eta\})$ .
 
 We consider a temperature which is much smaller than the energy
 difference between in-plane and out-of-plane vortices. After having
 been initially in
 the state $\{\eta_0\}$, the system reaches first a quasi-equilibrium 
 state near the
 metastable point $\{\eta_0\}$  with the probability  density ${\cal P}$ 
 given by the Gibbs distribution
 \begin{eqnarray}
 \label{hibbs}
 {\cal P} \sim e^{-\frac{E(\{\eta\})}{T}}~~~
 \mbox{with } \{\eta\} \simeq \{\eta_0\}.
 \end{eqnarray} 
 The probability flux over the barrier
 is concentrated in a narrow region around  the saddle point
 $\{\bar{\eta}\}$ \cite{htb90}.  To obtain the flux, let $D_{ni}, 
 (n,i=1,..2\,N)$ be the eigenvectors of the Hessian
 matrix 
 \begin{eqnarray}
 \label{hessian}
 {\cal E}_{i,j}(\{\eta\})=
\partial^2\,E/\partial\,\eta_i\,\partial \eta_j,~~~i,j=1,..2 N
 \end{eqnarray} 
 evaluated at $\{\eta\}=\{\bar{\eta}\}$
 \begin{eqnarray}
 \label{eq25}
 \sum_{j=1}^{2 N}{\cal E}_{i,j}(\{\bar{\eta}\})\,D_{lj}=\mu_l\,D_{li}
 \end{eqnarray}
  and $\mu_l$ are the eigenvalues. Thus the energy of the system in the
  immediate neighborhood of the saddle point $\{\bar{\eta}\}$ can be
  written as
 \begin{eqnarray}
 \label{eq26}
 E=E(\{\bar{\eta}\})+\frac{1}{2}\sum_{l=1}^{2 N}\,\mu_l\,\xi^2_l
 \end{eqnarray}
 where the new variables
 \begin{eqnarray}
 \label{eq27}
 \xi_{l}=\sum_{i=1}^{2 N}\,D_{li}\,(\eta_i-\bar{\eta}_i)
 \end{eqnarray}
  are the principal axes coordinates. 
  
  Coming back to the original variables $m_{\vec{n}}$ and
  $\phi_{\vec{n}}$ we can say that the Hamiltonian of the system in
  the close vicinity to the in-plane vortex state can be written as
  \begin{eqnarray}
 \label{eq28}
  \lefteqn{H=E_{\rm in-plane}}
  \nonumber\\
  &&\qd+\frac{1}{2}\,J\,\sum_{\vec{n},\vec{\Delta}}\left(\frac{1}{2}
  \cos(\Phi^0_{\vec{n}}-\Phi^0_{\vec{n}+\vec{\Delta}})\,(\psi_{\vec{n}}- 
\psi_{\vec{n}+\vec{\Delta}})^2+
\cos(\Phi^0_{\vec{n}}-\Phi^0_{\vec{n}+\vec{\Delta}})\,m^2_{\vec{n}}-\lambda\,
m_{\vec{n}}\,m_{\vec{n}+\vec{\Delta}}\right)
\end{eqnarray}
where $\psi_{\vec{n}}=\Phi_{\vec{n}}-\Phi^0_{\vec{n}}$ are small
deviations of the in-plane angles from their static values
$\Phi^0_{\vec{n}}$. The out-of-plane spin deviations $m_{\vec{n}}$ are
also assumed to be small. In this case the eigenvalues $\mu_{l}$
correspond to the linear spin-wave spectrum of the system in the
presence of an in-plane vortex.  The normal modes were investigated in
\cite{wysin95} and it was found out that there is a particular soft
mode (its frequency goes to zero for $\lambda\rightarrow \lambda_c,
\lambda\,\leq\,\lambda_c$) which is responsible for the crossover from
the in-plane to the out-of-plane vortex structure.  In the interval
$\lambda\,>\,\lambda_c$ this mode becomes unstable. In terms of Eq.
(\ref{eq25}) it means that the corresponding eigenvalue, say $\mu_1$,
is negative.

According to \cite{langer69} (see also \cite{htb90}) the rate constant
for an escape from the metastable point $\{\eta_0\}$ via the saddle
point $\{\bar{\eta}\}$ has the form
\begin{eqnarray}
 \label{eq29}
\kappa=\frac{|\nu|}{2\pi}\sqrt{\frac{\det\left({\bf {\cal E}}
(\{\eta_0\})/\sqrt{2\pi\,T}\right)}{\left|\det\left({\bf {\cal E}}
(\{\bar{\eta}\})/ \sqrt{2\pi\,T}\right)\right|}}\,
\exp\left(-\frac{E(\{\bar{\eta}\})-E(\{\eta_0\})}{T}\right)
\end{eqnarray}
where $|\nu|$ is the deterministic growth rate of the unstable mode
at the saddle point. The quantity $\nu$  is the negative eigenvalue of
the following eigenvalue problem
\begin{eqnarray}
\label{eq31}
\mu_l\,\sum_{l'=1}^{2 N}\tilde{M}_{l\,l'}\,U_{l'}=\nu\,U_l\enspace,
\end{eqnarray}
where $U_l$ are eigenvectors and
\begin{eqnarray}
\label{eq32}
\tilde{M}_{l\,l'}=\sum_{i,j=1}^{2 N}\,D_{l, i}M_{i, j}\,D_{l',j}\enspace .
\end{eqnarray}
Taking into account Eq. (\ref{eq25}) we can rewrite Eq. (\ref{eq32})
in the form
\begin{eqnarray}
\label{eq33}
\sum_{i,j=1}^{2 N}{\cal E}_{l,j}(\{\bar{\eta}\})\,M_{i,j}\,v_j=\nu\,v_i
\end{eqnarray}
where $v_i=\sum_{l=1}^{2 N}D_{l,i}U_l$.

Coming back to the original variables $\psi_{\vec{n}}$ and
$m_{\vec{n}}$ we obtain from Eqs. (\ref{eq24a}), (\ref{gamma}),
(\ref{anti}), (\ref{hessian}) that the switching rate between
out-of-plane vortices with opposite polarization is determined by the
expression
\begin{eqnarray}
\label{rate}
\kappa=\frac{|\nu|}{2\pi}\exp\left(-\frac{F_{in}-F_{out}}{T}\right)
\end{eqnarray}
where
\begin{eqnarray}
\label{free}
F_{out}=E_{out}+T\,\sum_m\,\ln\left(\omega_m({\rm out})/T\right)
\end{eqnarray}
is the free energy of the out-plane vortex and $\omega_m({\rm out})$ is the
m-th normal mode of the vortex.
\begin{eqnarray}
\label{free'}
F_{in}=E_{in}+T\,\sum_m'\,\ln\left(\omega_m(in)/T\right)+ 
T\,\ln\left(|\omega_1(in)|/T\right)
\end{eqnarray}
is an effective free energy of the in-plane vortex. In Eq. (\ref{free'})
the prime means the summation over the stable modes of the in-plane vortex and
$|\omega_1(in)|$ is the modulus of the purely imaginary frequency which
corresponds to the unstable mode of the in-plane vortex.

The deterministic growth rate $\nu$ is the negative eigenvalue of the 
eigenvalue problem
\begin{eqnarray}
\label{eigen}
\sum_{\vec{n}'}\left(\frac{\partial^2\,H}{\partial 
\Phi_{\vec{n}} 
\partial \Phi_{\vec{n}'}}\right)_{m_{\vec{n}}=0,\Phi^0_{\vec{n}}}
(\gamma v^{(1)}_{\vec{n}'}-v^{(2)}_{\vec{n}'})&=&
\nu v^{(1)}_{\vec{n}}\nonumber\\
\sum_{\vec{n}'}\left(\frac{\partial^2\,H}{\partial 
m_{\vec{n}} \partial m_{\vec{n}'}}\right)_{m_{\vec{n}}=0,\Phi^0_{\vec{n}}}
(\gamma v^{(2)}_{\vec{n}'}+v^{(1)}_{\vec{n}'})&=&
\nu v^{(2)}_{\vec{n}}
\end{eqnarray}
where $v^{(1)}_{\vec{n}},v^{(2)}_{\vec{n}'}$ are the components of the 
eigenvector and we took into account that in the vicinity of the
saddle point one can neglect the dependence on $m_{\vec{n}}$ 
in the damping constants $\Gamma_i$.

Let us evaluate these formulae in the crude approach already used in
section 1. We consider
the core dynamics taking into account only one pair of canonically
conjugated variables $m_1$ and $\psi_1$ and putting 
$m_2=m_3=0$, $\psi_2=\psi_3=0$ in Eq. (\ref{eq17}). In this case the 
eigenvalue problem  
(\ref{eigen})
reduces to
\begin{eqnarray}
\label{eq34}
\nu\,v^{(1)}-8J\lambda_c\,(\gamma v^{(1)}-v^{(2)})&=&0\enspace,
\nonumber\\
\nu\,v^{(2)}+8J(\lambda-\lambda_c)\,(\gamma v^{(2)}+v^{(1)})&=&0
\end{eqnarray}
where $\lambda_c=\cos\delta_1$. The deterministic
growth rate $\nu$ takes on the form 
\begin{eqnarray}
\label{eq35}
\nu=8J\left((\lambda_c-\frac{\lambda}{2})\gamma-
\frac{1}{2}\sqrt{\lambda^2\gamma^2+4(\lambda-\lambda_c)
\lambda_c}\right)\enspace.
\end{eqnarray}
The out-of-plane vortex exists for $\lambda\,>\,\lambda_c$ and 
the static value of the out-of-plane spin deviation is
$m^0_1=\pm\sqrt{1-\frac{\lambda_c^2}{\lambda^2}}$. The Hessian matrix
(\ref{hessian}) is evaluated at the metastable point (out-of-plane
vortex) and at the saddle point (in-plane vortex) which yields
  \begin{equation}4\,J\, \left(\begin{array}{clcr}
 \frac{\lambda\,(\lambda^2-\lambda_c^2)}{\lambda_c^2}&0\\ 
 0&\frac{\lambda_c^2}{\lambda}
 \end{array}\right),~~~
 4\,J\, \left(\begin{array}{clcr}
 -(\lambda-\lambda_c)&0\\ \label{eq36}
 0&\lambda_c
 \end{array}\right)\end{equation}
respectively. Inserting Eqs. (\ref{eq35}), (\ref{eq36}) into (\ref{eq29})
yields
\begin{eqnarray}
\label{eq37}
\kappa=2J\frac{\sqrt{\lambda^2\gamma^2+4\,\lambda_c\,(\lambda-
\lambda_c)}+\gamma\,(\lambda-2\lambda_c)}{\pi}\sqrt{\frac{\lambda+
\lambda_c}{\lambda_c}}\,e^{-\frac{E_{in}-E_{out}}{T}}
\end{eqnarray}
 We see that in the low-damping limit the
 switching rate reduces to
 \begin{eqnarray}
\label{eq38}
\kappa=\frac{2J}{\pi}\sqrt{\lambda^2-
\lambda_c^2}\,e^{-\frac{E_{in}-E_{out}}{T}}
\end{eqnarray}
while in the overdamped limit
\begin{eqnarray}
\label{eq39}
\kappa=\frac{2J\gamma}{\pi}(\lambda-\lambda_c)\sqrt{\frac{\lambda+
\lambda_c}{\lambda_c}}\,e^{-\frac{E_{in}-E_{out}}{T}}\enspace.
\end{eqnarray}
We note that the expressions (\ref{eq37}), (\ref{eq38}) and
(\ref{eq39}) are valid only when $\frac{E_{in}-E_{out}}{T}\gg
1$. This condition is not fulfilled when $\lambda\rightarrow \lambda_c$.

\section{Langevin dynamics simulations}

In order to test our theory we have numerically integrated the
stochastic Landau-Lifshitz equation (\ref{eq3}) for a large square
lattice in which we cut out a circle with radius $L$ using free
boundary conditions. As initial spin configuration we take an
out-of-plane vortex with center at a distance $R_0$ from the middle of
the circle. Since the anisotropy parameter $\lambda$ should not be
chosen close to $\lambda_c$ (see section II), the diameter $2r_v$ of
the out-of-plane vortex structure in any case is considerably larger
than the lattice constant. This has the advantage that the vortex can
move smoothly over the Peierls-Navarro potential of the lattice;
indeed discreteness effects are hardly visible in the motion.

Without noise and damping the trajectory $\vec{X}(t)$ of the vortex
center would be a circle with radius $R_0$ in a first approximation
which is given by the Thiele equation \cite{thiele73}
\begin{eqnarray}
\label{thiele_eq}
\vec{G}\times\dot{\vec{X}}=\vec{F} \enspace.
\end{eqnarray}
The driving force $\vec{F}$ is the 2D Coulomb force between the vortex
and an image vortex which is located at the distance $L^2/R_0$ from
the circle center \cite{Mertens94}. The image has opposite vorticity
but the same polarization as the vortex (for free boundary
conditions).  Eq. (\ref{thiele_eq}) was derived from the
Landau-Lifshitz equation in the continuum limit, assuming a rigid
vortex shape. In a better approximation the trajectories turn out to
be a superposition of cycloids around the circular motion
\cite{Mertens97}, but this fact seems to be unimportant for the
switching process which we discuss here.

When the damping term in (\ref{eq3}) is included, the vortex moves
outwards on a spiral \cite{Voelkel91}, until it finally reaches the
boundary where an annihilation together with the image takes place.
However, we choose an initial position far away from the boundary and
a very small damping parameter; therefore we have plenty of time to
observe the motion of the vortex before it gets close to the boundary.

When the stochastic fields $\vec{h}_{\vec{n}}(t)$ in (\ref{eq3}) are
included, the vortex trajectories naturally become noisy. In this case
the variances $\langle X_i^2 \rangle -\langle X_i \rangle^2$ can be
computed as a function of time and can be compared with a collective
variable theory for finite temperature \cite{Kamppeter97},
\cite{Kamppeter98}, \cite{Kamppeter98a}. This yields an effective
vortex diffusion constant $D_{\rm v}$.

In contrast to the vorticity $q$, the polarization $p$ of the vortex
is not a constant of motion in a discrete system: The out-of-plane
vortex structure can flip to the other polarization due to the stochastic
fields. Then the direction of $\vec{G}=2\pi q p \vec{e}_z$ is reversed
and thus the direction of the vortex motion is reversed, too, as can
be seen from (\ref{thiele_eq}).

In order to measure the transition rate $\kappa$ in the simulations it
is necessary to choose carefully the parameter ranges: $\lambda$ has
already been discussed above, we take $\lambda=0.9$ which is
sufficiently far away from both $\lambda_c \simeq 0.70$ and the
isotropic limit $\lambda=1$. For our circular system we choose a
radius $L=24$ which provides enough space for the vortex (the
out-of-plane vortex structure should not contact the boundary even
during long integration times). For the same reason the initial
distance $R_0$ of the vortex center from the middle of the circle
should not be too large. On the other hand $R_0$ should not be too
small, otherwise the driving force $\vec{F}$ would not be strong
enough to overcome the pinning forces of the lattice. Choosing $R_0
\simeq 10$ both conditions can be fulfilled, if the damping $\gamma$
is small enough. (The larger $\gamma$ is, the sooner the vortex reaches
the boundary). On the other hand, a small $\gamma$ means a long
saturation time (after the start of the simulation the energy rises
and saturates at a value independent of $\gamma$). For $\gamma \geq
0.002$ we get acceptable saturation times $<300$ (in units of
$\hbar/(JS)$).

The most important parameter naturally is the temperature: For $T \ll
E_{in} - E_{out}$ the transition rate $\kappa$ in (\ref{eq37}) is
extremely small and thus the integration times would be much too long,
which are needed to get a sufficient number of transition events. 

On the other hand $T$ should not be too large, otherwise
vortex-antivortex pairs appear spontaneously in the vicinity of the
vortex. This definitely changes the translational motion of the
vortex, and it is possible that the transition to the other
polarization is influenced, too. The difference $E_{in} - E_{out}$ can
be estimated by comparing the total energies of our system with $L=24$
in the presence of a static in-plane or out-of-plane vortex at the
center of a lattice cell: $E_{in} - E_{out} = 109.40 - 108.49 = 0.91$
(in units of $J$). The factor in front of the exponential in Eq.
(\ref{eq38}) is approximately $0.12$ therefore $0.1 \leq T \leq 0.3$
is expected to be an appropriate temperature range (The
Kosterlitz-Thouless transition temperature is about $0.8$ for
$\lambda=0$).

The initial spin configuration for our simulations stems from an
iterative program \cite{Schnitzer96} which produces a discrete vortex
structure on the lattice (In this way we avoid the radiation of spin
waves which would occur during the first time units if a continuum
approach for the vortex structure were used). As we interprete the
Landau-Lifshitz Eq. (\ref{eq3}) as a Stratonovich stochastic equation
and as we use multiplicative noise, we take the Heun integration
scheme which was developed for this situation \cite{Gard}, 
\cite{SanMiguel97}. The spin length $S$ is conserved in
Eq. (\ref{eq3}) and can be used as a test of the program, the time
step is $0.01$, in units of $\hbar/(JS)$.

For reasons to discuss below, we have performed two different
types of simulations: In type I a complete simulation for one
temperature consists of many runs with different sequences of random
numbers which produce the white noise. The total integration time is
divided into a first part of length $t_0$ (denoted as pre-run) and a
second part of length $t$ (denoted as main run). We choose $t_0$ in
the order of 1000 which is larger than the saturation time and large
enough that the vortex has no memory of the configuration from which it
started; i. e. in every run we have at the time $t_0$ a different
initial condition for the main run. Only the main runs are used for
the thermal average: the average time $\tau$, after which the {\it
  first} transition of the vortex to the opposite polarization occurs,
is obtained from
\begin{eqnarray}
\label{expdecay}
N(t)=N_0e^{-t/\tau}\,.
\end{eqnarray}
Here $N_0$ and $N$ are the number of runs in which the vortex has made
no transition until $t_0$ and $t_0 + t$, respectively. $\tau$ must be
compared with the inverse transition rate $\kappa^{-1}=\tau_{th}$
from Eq. (\ref{eq38}), because we work with a small damping parameter
$\gamma=0.002$ (Table \ref{table1}). The agreement is rather good,
taking into account that we used a very crude model for the vortex
core formed from only the four innermost spins.

We counted only the first transitions because in our theory we have
calculated the escape rate from a metastable state. After the first
transition the vortex is typically in a different dynamical state
than before, thus the probability for the next transition is expected
to be different, too. In fact, we obtained a total number of $870$
transitions in $158$ runs with $t=4000$ for $T=0.15$; this means that
the average transition time is 917, which is about four times smaller
than the first-transition time $4286$ in Table \ref{table1}.

In a type-II simulation we only make one pre-run of length $t_0$, i.
e. the main runs all start from the same initial condition. By taking
different lengths $t_0$ we can see whether $\tau$ depends on $t_0$
and/or the initial condition. We performed this type of simulations
because we had some hints from the investigation of the variances
\cite{Kamppeter98} that a certain vortex mode might be gradually
excited thermally which could trigger the transition. The frequency of
this mode is very low, namely $\Delta\omega = \omega_1-\omega_2$,
where $\omega_{1,2}$ are the eigenfrequencies of two quasi-local modes
of the circular system with one vortex \cite{Ivanov98}. $\omega_{1,2}$
are identical to the frequencies of the cycloidal oscillations of the
vortex trajectory around the mean path (see above).

However, our type-II simulations in Table \ref{table1} do not reveal a
correlation between the length $t_0$ of the pre-run and the first
transition time $\tau$. Nevertheless the values of $\tau$ differ
considerably for the different simulations. Thus we conclude that
$\tau$ depends strongly on the initial condition, which is identical
for all main runs of one simulation. This conclusion is confirmed by
looking at the first $200$ time units imediately after the beginning
of the main runs: E. g., in simulation No. 7 about $20\%$ of the
vortices switched over to the other polarization, while in No. 6 no
vortex did so (Fig. 3). A closer inspection of the initial spin
configurations shows that $\tau$ depends both on the position of the
vortex center within a lattice cell and on the dynamical state of the
vortex.

An additional test of the above conclusion was made by leaving out the
first $500$ time units of each main run (dash-dotted line in Fig. 3).
Then we expect that the vortices have no memory of their initial
condition and the resulting $\tau$ should be the same as in the type-I
simulations (within the statistical errors). In fact, this is
confirmed by comparing No. 8 with No. 2 in Table I.


\section{Conclusion}

In this work we used a very simplified Hamiltonian for the cores of
both planar and nonplanar vortices. Adding white noise to the local
fields in which the classical spins precess we obtained a
Landau-Lifshitz equation with multiplicative stochastic forces. The
stationary solution of the corresponding Fokker-Planck equation
exhibits two maxima for the two possible polarizations of the
nonplanar vortex and a saddle point for the planar vortex, if the
anisotropy parameter lies in a certain, temperature-dependent range.

We calculated the rate $\kappa$ for the transition from one
polarization to the opposite one. Our results were tested by long-time
Langevin dynamics simulations of the full many-spin model at three
temperatures well below the Kosterlitz-Thouless phase transition
temperature. The agreement is rather good, considering that the vortex
core was described only approximately by using only the four innermost
spins. We did not make any tests for higher temperatures because the
probability for the spontaneous appearance of a vortex-antivortex pair
in the vicinity becomes too large; the interaction with this pair
could then influence $\kappa$.

We emphasize that the above results were obtained by effectively
averaging over many initial conditions. This is necessary because our
simulations demonstrate that the transition rate depends very strongly
on the initial condition, i. e. both on the position of the vortex
center within a lattice cell and on the velocity of the vortex at this
position.

\section{Acknowledgements}
Yu. Gaididei would like to express 
his thanks for the hospitality of the University of Bayreuth 
where this work was done. Partial support was received
from project No. X 271.5 of the scientific and technological
cooperation between Germany and Ukraine. Work at Los Alamos National 
Laborartory is supported by the USDoE.
%

%
\clearpage

\centerline{\epsfysize=20.0truecm \epsffile{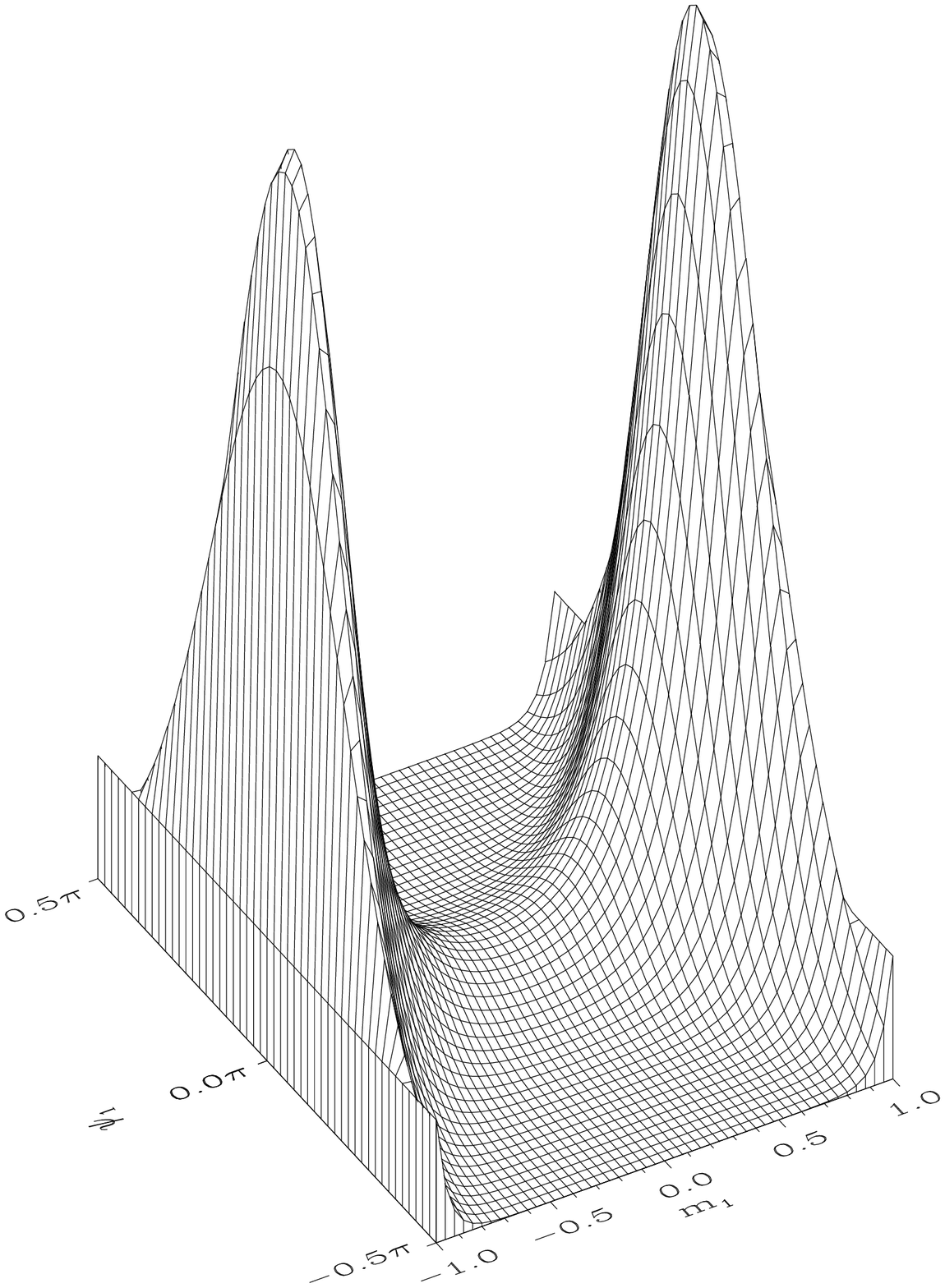}} Fig. 1. (a)
Stationary probability distribution (14) (without normalization),
using the simplified core Hamiltonian (26). $m_1$ and $\psi_1$ are the
deviations (20) and (19) of the out-of-plane components and in-plane
angles from their static values, resp..\\

\centerline{\epsfysize=10.0truecm \epsffile{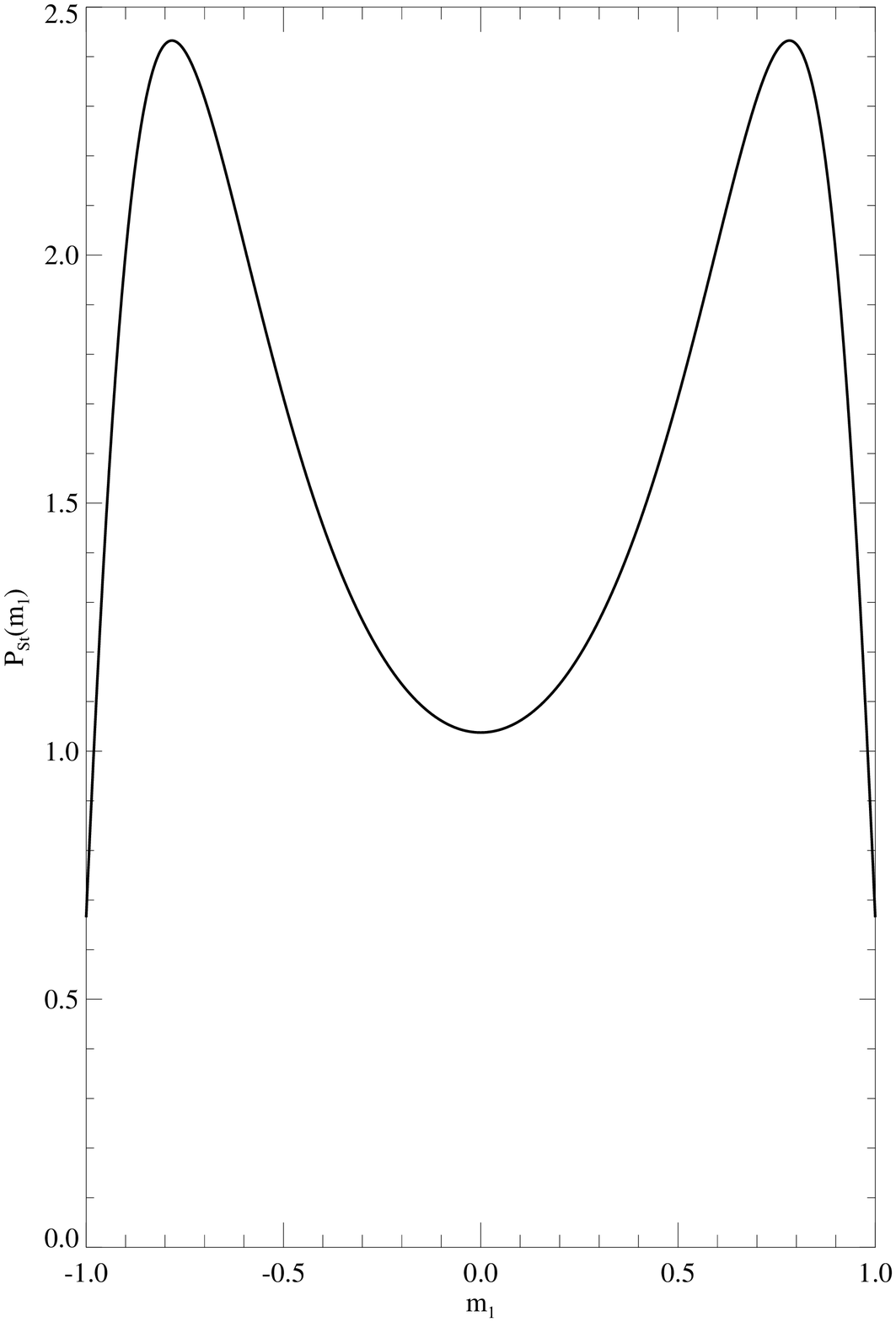}} (b) Reduced
stationary distribution (27), obtained by integrating the
distribution in $(a)$ over $\psi_1$.\\

\centerline{\epsfysize=7.0truecm \epsffile{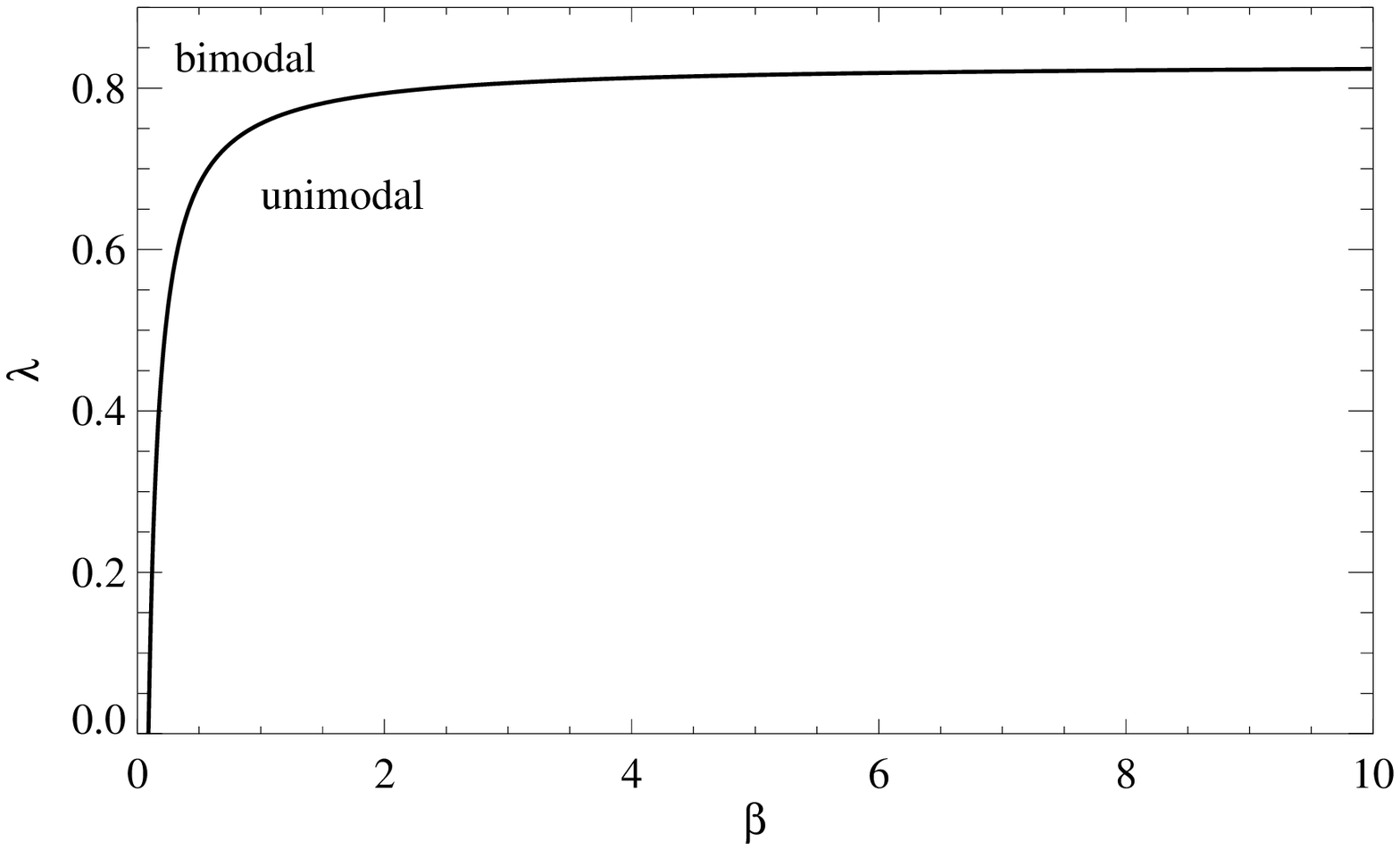}} Fig. 2. $\lambda$
vs. $\beta$ phase diagram. Above the bifurcation curve
$\lambda(\beta)$ there are two maxima in ${\cal P}_{st}$,
corresponding to the two polarizations of a nonplanar vortex, and one
saddle point corresponding to
a planar vortex structure.\\

\centerline{\epsfysize=7.0truecm \epsffile{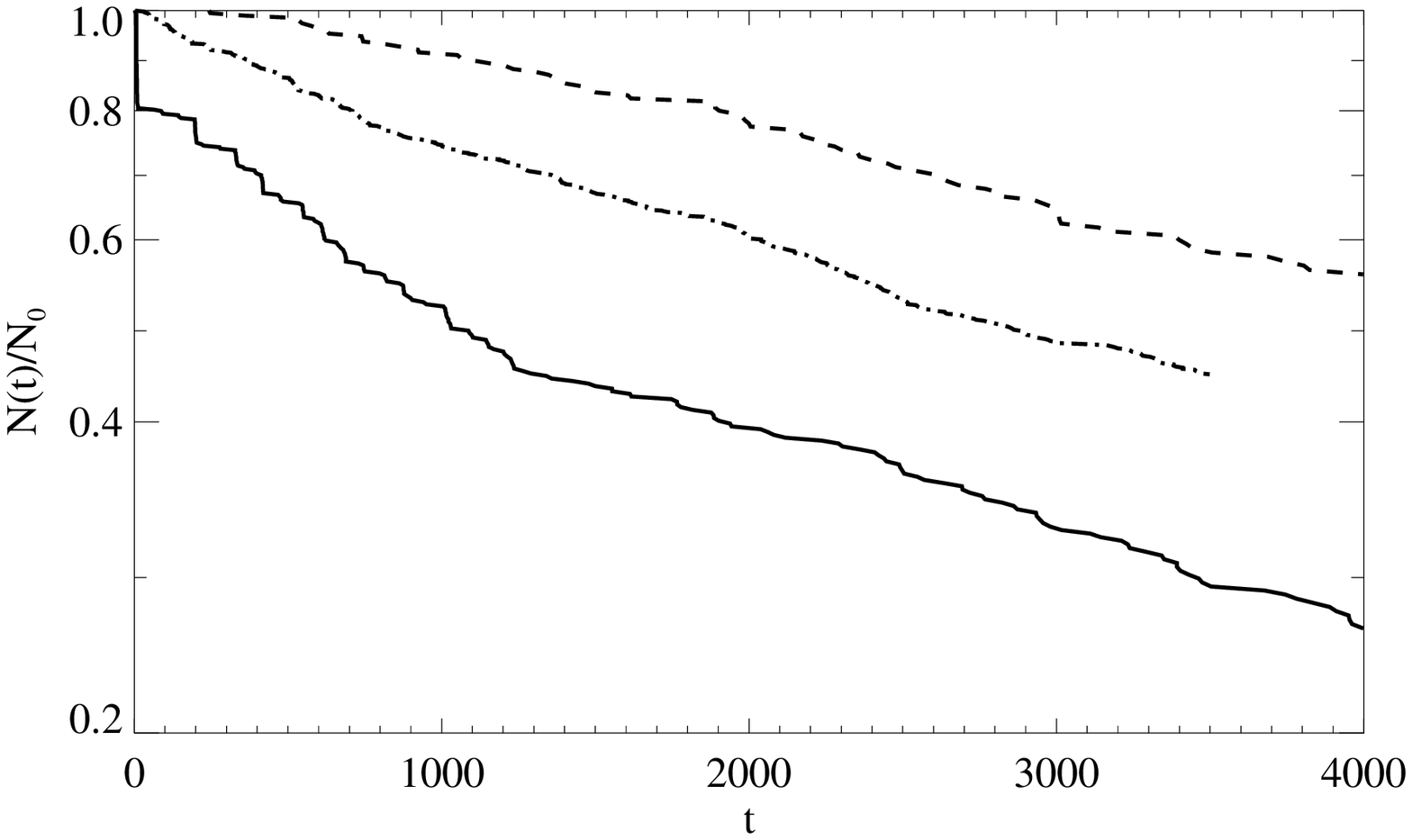}} Fig. 3.
Percentage of vortices which have not yet made a transition to the
opposite polarization up to the time $t$ of a type-II simulation. The
temperature is $T = 0.15$. The solid and dashed lines represent two
sets of runs (No. 7 and 6 in Table I) with different initial
configurations (which arise from using two different lengths $t_0$ for
the pre-run). The dash-dotted line results from sampling the
transition times from the simulations No. 4 - 7 of Table I,
omitting the first 500 time units of each main run.\\
 
\begin{table}
\begin{center}
\caption{Transition times $\tau$ from simulations, with statistical
  errors $\tau_{\rm rms}/\tau$, compared to the theoretical estimates
  $\tau_{\rm th}$\\}
\label{table1}
\begin{tabular}{cccccccccc}
No. & type & $T$ & $t_0$ & $t$ & 
$N_0$ & $N$ & $\tau$ & $\tau_{\rm rms}/\tau$ & $\tau_{\rm th}$ \\\hline
1 & I & 0.1 & 1200 & 3800 & 497 & 477 & 92516 & 22\% & 70334 \\
2 & I & 0.15 & 1200 & 3800 & 407 & 158 & 4016 & 6\% & 3386 \\
3 & I & 0.2 & 1200 & 3800 & 100 & 1 & 825 & 10\% & 743 \\\hline
4 & II & 0.15 & 1200 & 4000 & 254 & 100 & 4291 & 8\% & -- \\
5 & II & 0.15 & 2200 & 4000 & 264 & 100 & 4120 & 8\% & -- \\
6 & II & 0.15 & 3200 & 4000 & 181 & 100 & 6741 & 11\% & -- \\
7 & II & 0.15 & 4200 & 4000 & 405 & 100 & 2859 & 6\% & -- \\\hline
8 & II & 0.15 & -- & 3500 & 905 & 400 & 4286 & 5\% & 3386 \\
\end{tabular} 
\end{center}
\end{table}
\end{document}